# Identification of complex mixtures for Raman spectroscopy using a novel scheme based on a new multi-label deep neural network

Liangrui Pan, *Member, IEEE*, Pronthep Pipitsunthonsan, Member, IEEE, Chalongrat Daengngam, Mitchai Chongcheawchamnan, Senior, Member, IEEE

*Abstrghact*— With noisy environment caused by fluoresence and additive white noise as well as complicated spectrum fingerprints, the identification of complex mixture materials remains a major challenge in Raman spectroscopy application. In this paper, we propose a new scheme based on a constant wavelet transform (CWT) and a deep network for classifying complex mixture. The scheme first transforms the noisy Raman spectrum to a two-dimensional scale map using CWT. A multi-label deep neural network model (MDNN) is then applied for classifying material. The proposed model accelerates the feature extraction and expands the feature graph using the global averaging pooling layer. The *Sigmoid* function is implemented in the last layer of the model. The MDNN model was trained, validated and tested with data collected from the samples prepared from substances in palm oil. During training and validating process, data augmentation is applied to overcome the imbalance of data and enrich the diversity of Raman spectra. From the test results, it is found that the MDNN model outperforms previously proposed deep neural network models in terms of Hamming loss, one error, coverage, ranking loss, average precision, F1 macro averaging and F1 micro averaging, respectively. The average detection time obtained from our model is 5.31 s, which is much faster than the detection time of the previously proposed models.

## I. INTRODUCTION

Raman spectroscopy is a fast, non-invasive, label-free and no pretreatment technology, which can display molecular fingerprints according to vibration information [1]. Since Raman spectroscopy is insensitive to water, hence it has been widely used in several applications such as chemistry [2], materials [3], physics [4], polymer [5], biology [6], medicine [7] and geology [8]. Identification of organic chemistry using Raman spectroscopy is achieved by an interaction of molecular structure with infrared spectrum. The spectrum_characteristics, which are magnitudes of Raman shift, peak intensity and peak shape, are the vital basis for identifying chemical bonds and functional groups. There have been several research works on Raman spectroscopy technology; for example surface-enhanced Raman spectroscopy [9], high-temperature Raman spectroscopy [10], resonance Raman spectroscopy [11], confocal micro Raman spectroscopy [12], Fourier transform Raman spectroscopy [13], to name a few. These techniques promote the application of Raman spectroscopy in various fields.

One of the disadvantages of Raman spectroscopy is that it is easily interfered by fluorescence noise. Once the sample under test responds with fluorescence, the Raman spectrum will be swamped by wideband spectrum of fluorescence noise. This causes the desired Raman spectrum hardly to be detected. Secondly, the sensitivity of Raman spectroscopy is low. There are several unavoidable noise such as shot noise, dark current noise and readout noise [14], [15] in a Raman detector implemented with charge coupled devices (CCD) and semiconductor devices. Therefore, before using Raman spectroscopy, a preprocess algorithm such as baseline correction is needed to reduce the interference of these noises and highlight the molecular peak characteristics. Two baseline correction approaches which are based on hardware and software designs have been proposed [16]–[18]. The hardware design approach needs an instrument modification, hence, is unpopular. The software design approach, on the other hand, is based on signal processing technique. With no additional hardware installation and modification, it is the low cost approach and thus gains more interest.

In recent decades, some researchers proposed using Raman spectroscopy to promote the research progress in other fields. Corey *et al.* demonstrated a multi-stage rapid classification method for minerals in Raman spectroscopy using an extensive RRUFF database [19]. Lianrui *et al.* proposed the deep neural network (DNN) to classify the Raman spectrum of hazardous chemicals. The obtained classification accuracy reaches 99% [20]. Xiangxiang *et al.* proposed a rapid but low-cost method to detect thyroid dysfunction using serum Raman spectroscopy and support vector machine (SVM) [21]. Jing *et al.* proposed a method based on confocal Raman spectroscopy and SVM to distinguish lung adenocarcinoma cells from normal cells [22]. Based on the surface-enhanced Raman spectroscopy of bacterial samples, Geoffrey *et al.* used a measurement system including a portable low-cost Raman spectrum acquisition unit and signal processing as well as classification modules to distinguish six bacteria in Listeria [23].

Usually, using Raman spectroscopy to identify single pure material can be done. However most substances in nature is a mixture and it is rather challenging to identify materials in the mixture from the combined Raman spectrum peaks. Due to the rapid development of artificial intelligence, some models that stand out in the competition such as Visual Geometry Group with 16 layers (VGG16) [24], Visual Geometry Group with 19 layers (VGG19) [25], the 50-layer Residual network (ResNet50) [26], MobileNetV2 [27], DenseNet121 [28], and InceptionResNetV2 [29] can successfully migrate and being applied to specific scenarios. Based on the topology of these networks, some researchers proposed new algorithms to classify multi-label data.

Several multi-label classification methods were proposed [30]-[38]. Empirical research works on benchmarking multi-label classification applied to image data sets demonstrated that the performance of a deep multi-modal CNN for multi-instance multi-label in multi-label image classification tasks was significantly better than the previously proposed methods [30]. Bingzhi *et al.* proposed a new label co-occurrence learning framework based on graph convolution networks (GCNs) to clearly explore the pathological dependence in multi-label chest X-ray image classification tasks. Extensive experiments on ChestX-Ray14 and CheXpert data sets proved the effectiveness of GCNs as compared with the previouslt proposed methods [31]. Yuansheng *et al.* proposed a new multi-label classification network for aerial images, namely attention-aware label relational reasoning network [32]. In [33], Xin *et al.* proposes a new algorithm based on multi-label integration for complex remote sensing scene data. Ricardo *et al.* presented Hierarchical Multilabel Classification with a Genetic Algorithm (HMC-GA). It is a genetic algorithm for classification rule induction in hierarchical multi-label scenarios. They compared their proposed algorithm with three decision tree induction algorithms based on predictive clustering trees. [34] .

Based on the development limitations of multi-label classification network and algorithm, Yang *et al.* proposed a SVM based method for multi-label learning with missing label problems and solve the optimization problem through an iteratively re-weighted least squares (IRWLS) method [35]. Karl *et al.* proposed a novel semi-supervised and multi-label dimensionality reduction method, which effectively utilizes the information in noisy multi-label and unlabeled data. Experimenting with a large number of synthetic data and benchmark data sets, it was shown that the algorithm is effective and superior to the latest multi-label feature extraction algorithm [36]. By learning the high-order tag correlation, Jun *et al.* extends the inner complete tag matrix to a new complementary tag matrix. Then, the specific label data representation of each class tag is learned, and on this basis, combined with the learned high-order tag correlation. It was demonstrated that a multi-label classifier was constructed [37].

In Raman spectrum recognition for a mixture, it is needed to extract the molecular information of the mixture, that is, the Raman intensity corresponding to the Raman shift. The baseline correction algorithm is used to preprocess the Raman spectrum of the mixture, and different detection algorithms are used to determine the composition of the molecular fingerprint. This method requires high accuracy of Raman spectroscopy and has limitations in practical operation. Recently, Xiaqiong *et al.* proposed a novel approach entitled deep learning -based component identification (DeepCID), and established a convolutional neural network (CNN) model to predict the presence of components in mixtures [38]. In [39], rapid recognition of mixtures in complex environments was realized by establishing a fast Raman analysis model based on deep learning through data training, self-learning, and parameter optimization. Although Raman spectroscopy combined with DNN can facilitate to identify components in a mixture, accuracy of these abovementioned methods is still insufficient. In addition, their proposed methods has yet some limitations under noisy environment.

This paper proposes a multi-label based DNN algorithm for classifying Raman spectrum of a complex mixture. The proposed preprocessing method based on the wavelet transform can capture the molecular signature information from noisy Raman spectrum. To improve detection accuracy of the model, we propose a DNN based multi-label classification . The detail of feature extraction, feature mapping and label classification will be explained. The main contributions of this paper are as follows:

(1) The continuous wavelet transform (CWT) is proposed for preprocessing and its performance for spectrum extraction is justified by comparing with those of conventional methods such as short-time Fourier transform (STFT) and Wigner–Ville distribution (WVD). CWT is applied to decompose desired molecular information and noise from the noisy Raman spectrum.

(2) A new multi-label deep neural network (MDNN) algorithm model is proposed. The new model avoids overfitting and underfitting problems during the model development. Detection accuracy and detection time of the proposed MDNN will be compared with previously proposed models such as VGG16, VGG19, ResNet50, DenseNet121, InceptionResNetV2 and MobileNet50.

## II. MATERIALS AND METHODS

This section mainly describes the experimental data collection, data preprocessing method and simplified version of MDNN algorithm. Fig. 1 overviews the development of workflow of the proposed scheme. It consists of data collection, spectrum preprocessing, developing and testing the MDNN model.

### A. Data Collection

The data collection process is shown in Fig. 1. We set up the experiment for collecing spectra in the temperature-controlled room at 28°C. A dark room was set up such that no any light interfered during our measurement. BIM-6002a Raman spectrometer was used to collect the Raman spectra. From the specfications of the spectrometer, the signal-to-noise ratio (SNR) of the channel is 600:1 and the laser wavelength is 785nm.

Several complex mixture samples were prepared. In this paper, three kinds of chemical substances related to palm oil which are Oleic acid, Palmitic acid and Retinyl Palmitate were prepared for different proportions. At room temperature, Oleic acid and Retinyl Palmitate are solvent while Palmitic acid is solute. Oleic acid, Retinyl Palmitate and Palmitic acid were mixed at a ratio of 2:1:1 to prepare four kinds of mixtures. Heat was continuously applied to the mixture samples for 5 minutes such that the temerature was stabilized at 50°C. This guarantees that the solute and solvent have fused completely. After heating, the mixture

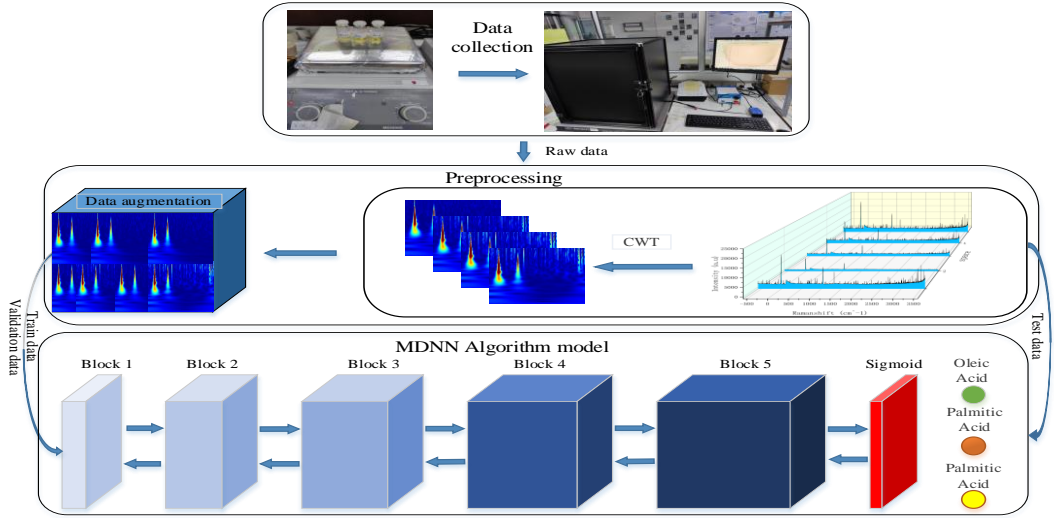

Fig. 1. Development flowchart of the proposed scheme .

was then placed in the Raman spectrometer and the spectra were measured and collected.

### B. Data Preprocessing- CWT analysis

To preprocess noisy Raman signal to denoise and highlight the molecular spectrum, several researchers commonly use the baseline correction method to remove the fluorescence noise. Though many baseline correction algorithms were proposed, these algorithms can denoise within a certain level [16]–[18], [40]. In this paper, we overcome this by transforming one-dimensional (1-D) noisy Raman septcrum to a two-dimensional (2-D) signal. CWT is proposed and its performances are investigated and compared with other 1-D to 2-D transformation algorithms, i.e. STFT and WVD.

The STFT lies under the assumption that the signal $x(t)$ is stationary in a short time controlled by the window function $g(t)$. The analysis is further assumed that $x(t)g(t-\tau)$ is also stationary in different finite time widths. Power spectrum at short-time window for different time snapshots can be calculated by [41]:

$$STFT(f,\tau) = \int_{-\infty}^{\infty} [x(t)g(t-\tau)]e^{-j2\pi ft}dt \quad (1).$$

It is shown in (1) that the window function divides the original signal into many time periods. The function of each time segment performs Fourier transform on it. The length of the window determines the time resolution and frequency resolution of the spectrum. The longer the window length is, the longer the intercepted signal is. The higher the frequency resolution is, the worse the time resolution is. On the contrary, the shorter the window length is, the shorter the intercepted signal is, the worse the frequency resolution is, and the better the time resolution is. There is a frequency-time resolution trade-off in STFT.

WVD is nautrally a typical quadratic transformation, which is defined as the Fourier transform of the signal instantaneous correlation function. Hence WVD reflects the instantaneous time-frequency relationship of the signal. Theoretically, WVD provides the best energy concentration and has many ideal mathematical characteristics [42]. Due to the quadratic transformation itself, the cross terms are generated for multi-component nonstationary signals. For any single component linear frequency modulated (LFM) signal, the projection of WVD on the time-frequency plane is a straight line, that is, the linear change of frequency with time. WVD is defined as:

$$W_S(t,f) = \int_{-\infty}^{\infty} s(t+\frac{\tau}{2})s^*(t-\frac{\tau}{2})e^{-j2\pi f\tau}d\tau \quad (2),$$

where $s(t+\frac{\tau}{2})s^*(t-\frac{\tau}{2})$ is the instantaneous autocorrelation function $R(t,\tau)$ of the signal $s(t)$. Since there is no window operation in the calculation, it avoids the mutual restraint between the time- and frequency -resolutions. For the single component LFM signal, the time-frequency representation of WVD has the best energy concentration. However, because it does not involve the window function, WVD will be interfered by cross terms when analyzing multi-component signals [42]. Some improved methods can suppress the influence of cross terms on WVD to some extent, but they can not eliminate the mutual interference of self cross term and multi-component of nonlinear FM signal at the same time.

CWT, on the other hand, expands the function $f(t)$ of any $L_2(R)$ space under the wavelet basis, which is defined as:

$$W_\varphi f(a,b) = |a|^{-\frac{1}{2}} \int_{-\infty}^{\infty} f(t)\varphi(\frac{t-b}{a})dt \quad (3),$$

where $\varphi(a,b)$ is the wavelet function, $a$ is a frequency scaling, parameter, and $b$ is a time-shifting parameter. Although STFT has time-shifting property, it is only suitable for stationary signals with small frequency fluctuation because the window size is fixed. WVD has better resolution than STFT, but there are cross interference terms. Compared with STFT and WVD, CWT has high sensitivity in processing abrupt signal, which is suitable for processing non-stationary signal similar to Raman spectrum, and shows high resolution without interference. In [43], experiments have proved the feasibility of using CWT to transform all the original Raman spectrum signals into 2-D multiresolution scale map for classification tasks.

STFT, WVD and CWT were applied to transform 1-D noisy Raman spectrum to 2-D scale map data. The noisy Raman spectrum of the mixture of Oleic acid and Retinyl Palmitate is shown in Fig. 2a. Transformation results with three algorithms are shown in Fig. 2b-d. As shown in Fig. 2b, STFT is not suitable for processing Raman spectrum with fluorescence noise and peak characteristics. In Fig. 2c, 2-D signal after applying WVD is shown. The molecular information and noise on 2-D data are not converted effectively. On the other hand, the result obtained from CWT in Fig. 2d shows that the noise and spectrum peak in 2-D scale map are well separated. The noise in the spectrum is shown as the white spot area where the spectrum peaks appear in the other color areas. This is suggested that CWT is superior to STFT and WVD for multiresolution analysis.

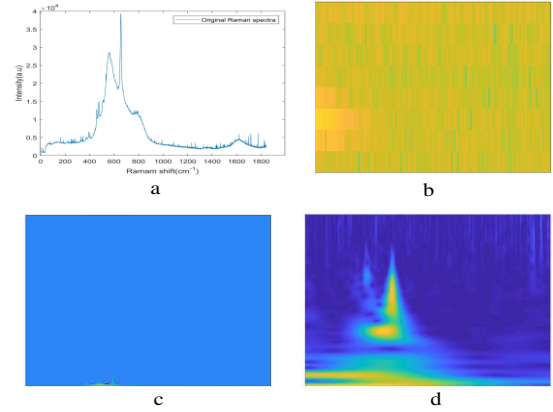

Fig. 2. STFT, WVD, CWT multiresolution analysis.

### C. Data Preprocessing- Data augmentation

Due to the difference in the sampling frequency and the power of the excitation light source, the original Raman spectroscopy data will be different in quantity. This problem can lead to overfitting during the learning stage of the model development. The synthetic minority oversampling technique (SMOTE) was proposed to overcome this problem [44]. Its concept is to analyze the minority samples and synthesize new samples according to the minority samples and add them to the data set. By adding a lot of different levels of noise to the raw data of pure substance and a small amount of noise of different levels to the raw data of mixture, the SNR which is used to process the signal is computed from [45]:

$$SNR(dB) = 10\log_{10}\frac{P_{signal}}{P_{noise}} \quad (4),$$

where $P_{signal}$ is the power of the input signal and $P_{noise}$ is the power of the input noise. We put the original data in the noise environment of 30-60 dB noise power. The noise signal is generated with random intensity. When each type of data in the data set is the same, data augmentation was performed. In the experiment, we choose the height and width range from 0 to 0.1. Some Raman spectrum scale were rotated by 90 degrees, horizontally or tilted, and randomly scramble all labels and data sets to prevent overfitting. As shown in Table I, the original Raman spectrum data and the number of data sets after data expansion are counted. The number of original Raman spectra ranged from 28 to 42. In order to ensure the balance of data in the experiment, the number of each spectrum is increased to 360. In the experiment, the ratio of training set and verification set is 8:2.

TABLE I   STATISTICS OF MIXTURE RAMAN SPECTRUM DATA SET.

| Material | Raw data | Augmented data |
|---|---|---|
| Oleic acid | 35 | 360 |
| Palmitic acid | 36 | 360 |
| Retinyl Palmitate | 28 | 360 |
| Oleic acid + Palmitic acid | 42 | 360 |
| Palmitic acid + Retinyl Palmitate | 28 | 360 |
| Oleic acid + Retinyl Palmitate | 28 | 360 |
| Palmitic acid + Oleic acid + Retinyl Palmitate | 28 | 360 |

### D. Method

A new multi-label deep neural network (MDNN) classification model is proposed. The model consists of six modules as shown in Fig. 3. Module 1-5 consists of a convolution layer and a pooling layer where module 6 consists of a global average pooling layer and multi-layer neural network. Module 7 is composed of *Sigmoid* layer and the classification task occurs here.

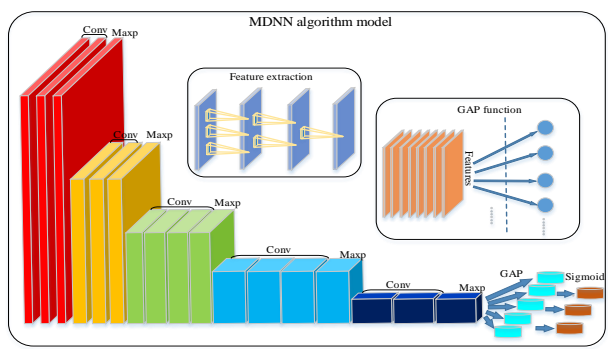

Fig. 3. MDNN model classification framework.

In Fig. 3, the model begins with two convolution layers and one pooling layer to extract a wide range of low-level features from the input image. Since feature extraction is a key process in multi-label classification algorithm, it is usually processed by using local sensing, parameter sharing, spatial sub-sampling and so on. For CNN, it is difficult and unstable to train the classification model (image to label) from randomly initialized convolution kernel. So it is necessary to transfer the parameters of multi-label classification model to the classification model. Therefore the combination of convolution layer and pooling layer in different number in the model is used. Through forward and backward propagation features, the model circulates and updates all learning parameters. The convolution kernel of $3\times3$ provides the maximum perception in the local field of view, and fuses the local sensing information at the highest level to obtain the global perception. In local sensing, the same features obtained by the filter can be shared with other filters to improve the efficiency of feature extraction. The main function of the $2\times2$ pooling window is to downsample the global information and reduce the calculation of parameters. Secondly, the nonlinear mapping of *Relu* function can expand the receptive field to realize translation invariance, rotation invariance and scale invariance [46]. The process of Raman spectrum from two convolution and one maximum pooling layers in the first part to four convolution layers and one maximum pooling laye in the fifth part is called feature extraction [47].

After feature extraction, all features are reshaped into vectors by using the full connection layer. Then the vectors are multiplied to reduce their dimensions following with the *Softmax* layer which is used for output [48]. This method not only changes the network parameters, but also causes overfitting problem. However, the global pooling layer is considered as a new technique to replace the full connection layer, which has a great effect on reducing parameters and reducing the risk of overfitting. In the experiment, the global average pooling is used instead of the full connection layer, which directly averages the entire feature map, and then input it into the *Sigmoid* layer to get the probability of tags and mappings [48]. By replacing the black box operation of the full connection layer, the network parameters are significantly reduced, hence avoiding overfitting problem.

### III. RESULTS

#### A. Evaluation index

Multi-label learning is to map a sample and a set of tags to an instance. Suppose $\chi = R^d$ is $d$ sample space and $\gamma = \{y_1, y_2, ..., y_q\}$ is label space. The task of multi-label learning is to learn a function $h : x = 2^Y$ from the training set $D = \{(x_i, Y_i) | \leq i \leq m\}$. For each multi-label sample $(x_i, Y_i)$, $x_i \in \chi$ is the vector of $d$-dimensional features, and $Y_i \in \gamma$ is the sum of label sets. For an unknown sample $x \in \chi$, the multi-label classifier $h(\cdot)$ predicts $h(x) \subseteq \gamma$ as the label set of the sample [49].

In the traditional supervised learning, the generalization performance of the algorithm model is evaluated by the accuracy, F-measure, ROC (AUC) and other traditional indicators. However, the performance evaluation in multi-label learning is much more complicated than the traditional single-label learning. Each example can be associated with multiple tags at the same time. Therefore, we use eight commonly used evaluation indicators in multi-label learning, namely Hamming loss, ranking loss, coverage, one-error, average precision, F1 micro averaging, F1 macro averaging and receiver operating characteristic (ROC) [50].

In label-based indicators, for j$^{th}$ class label $y_i$, four basic quantities that characterize the two-class classification performance of the label can be defined by:

$$\begin{aligned} TP_j &= |\{x_i \mid y_j \in Y_i \land y_j \in h(x_i), 1 \leq i \leq p\}|; \\ FP_j &= |\{x_i \mid y_j \notin Y_i \land y_j \in h(x_i), 1 \leq i \leq p\}|; \\ TN_j &= |\{x_i \mid y_j \notin Y_i \land y_j \notin h(x_i), 1 \leq i \leq p\}|; \\ FN_j &= |\{x_i \mid y_j \in Y_i \land y_j \notin h(x_i), 1 \leq i \leq p\}|. \end{aligned} \quad (5).$$

In addition, $TP_j$, $FP_j$, $TN_j$, $FN_j$ denote true, false positive, true negative, and false negative. It can be proved from (5) that, $TP_j + FP_j + TN_j + FN_j = p$, which is constant [51]. Based on the above four quantities, most of the classification measures of binary classification problems can be calculated and processed. Let $B(TP_j, FP_j, TN_j, FN_j)$ denotes the binary classification matrix, the label based classification measures are defined by [51]:

(1) Macro-averaging:
$$B_{macro}(h) = \frac{1}{q} \sum_{j=1}^{q} B(TP_j, FP_j, TN_j, FN_j) \quad (6)$$

(2) Micro-averaging:
$$B_{micro}(h) = B(\sum_{j=1}^{q} TP_j, \sum_{j=1}^{q} FP_j, \sum_{j=1}^{q} TN_j, \sum_{j=1}^{q} FN_j) \quad (7)$$

In the case-based indicators, four classification measures can be defined as follows:

(1) *Hamming loss* evaluates the number of times instance tags which are misclassified. Predicting tags that do not belong to an instance or do not predict tags that belong to the instance are counted.

$$hloss(h) = \frac{1}{p} \sum_{i=1}^{p} \frac{1}{q} | h(x_i) \triangle Y_i | \quad (8),$$

where $\triangle$ is the symmetry difference between two sets. It is noted that for all instances $|Y_i| = 1$. A multi-label system is actually a multi-class single label system, while Hamming loss is usually 2/Q times of the classification error.

(2) *One-error* calculates the proportion of instances where the top-level tags are not in the set of related tags. One-error can be intepreted as the score of evaluating the reverse tag pair.

$$one - error = \frac{1}{p} \sum_{i=1}^{p} \{ [\arg\max_{y \in \gamma}] \in Y_i \} \quad (9).$$

(3) *Coverage* expresses how far, on average, we need to move down the rank list of labels so as to cover whole ground true labels of the object. The smaller value of coverage the better performance of algorithm. [52].

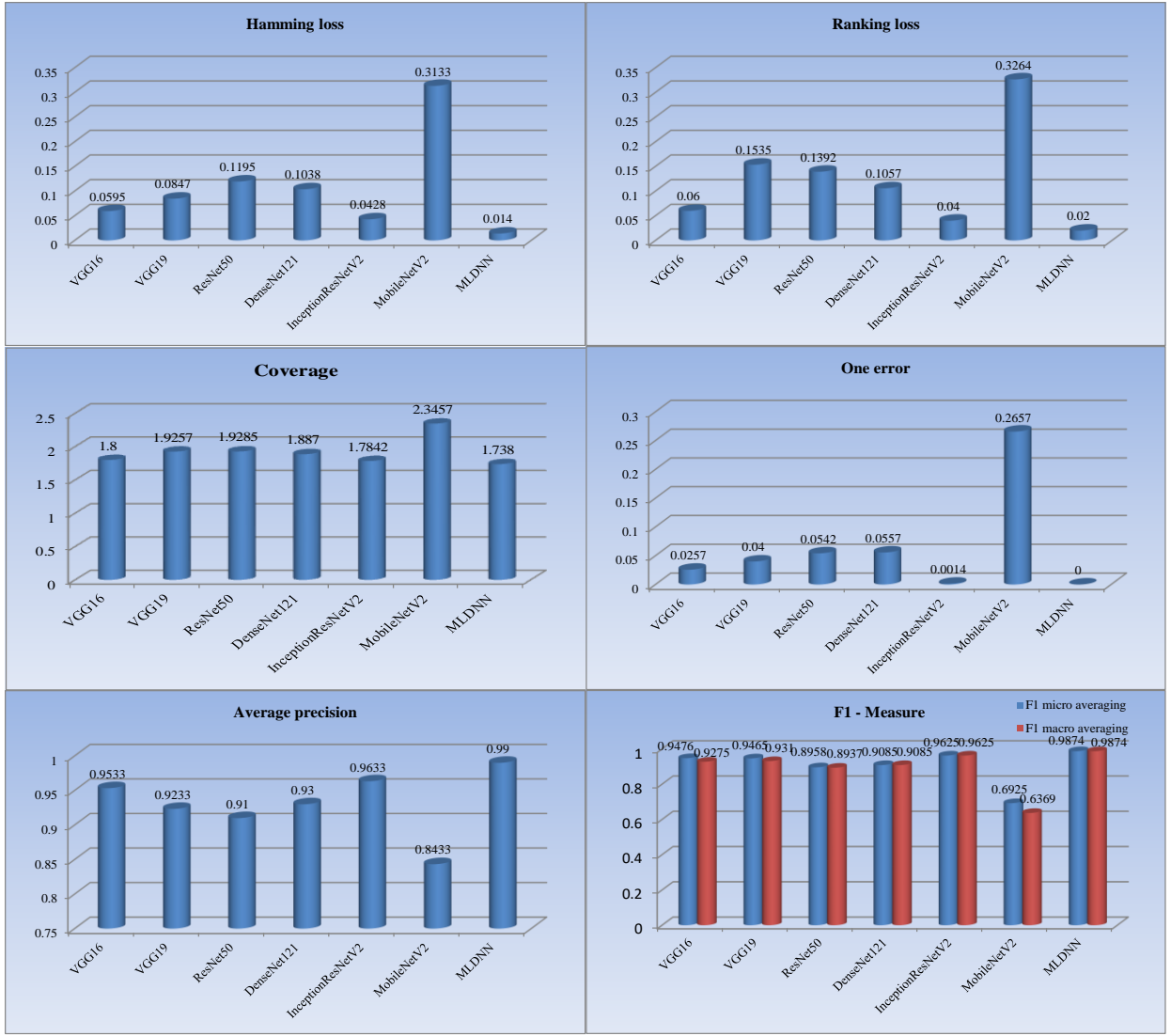

Fig. 4. Analysis of VGG16, VGG19, DenseNet121, InceptionResNetV2, MobileNetV2 and MDNN models on Hamming loss, one error, coverage, ranking loss, average precision, F1 macro averaging and F1 micro averaging.

$$\mathrm{cov}erage(f) = \frac{1}{p}\sum_{i=1}^{p}(\max_{y \in Y_i} rank_f(x_i, y) - 1)$$

(10)(4) *Ranking loss* calculates the average percentage of wrong instruction label pairs, that is, the irrelevant labels of an object is ranked higher than relevant labels. [52].

$$rloss(f) = \frac{1}{p}\sum_{i=1}^{p}\frac{1}{|Y_i||\overline{Y_i}|}|\{(y', y'') | f(x_i, y') \le f(x_i, y''), (y', y'') \in Y_i \times \overline{Y_i}\}|$$ (11).

(5) Average precision calculates the average score, including the actual tags. This is defined from:

$$avgprec_s(f) = \frac{1}{p}\sum \frac{1}{|Y_i|} \times \sum_{y \in Y_i} \frac{|\{y^i | rank_f(x_i, y') \le rank_f(x_i, y), y' \in Y_i\}|}{rank_f(x_i, y)}$$ (12).

For Hamming loss, one error, coverage and ranking loss, the smaller the value, the better the performance of the algorithm model. For average precision, the larger the measure, the better the model. The best values of Hamming loss, one-error, coverage, and ranking loss are 0 where the best value of average precision is 1.

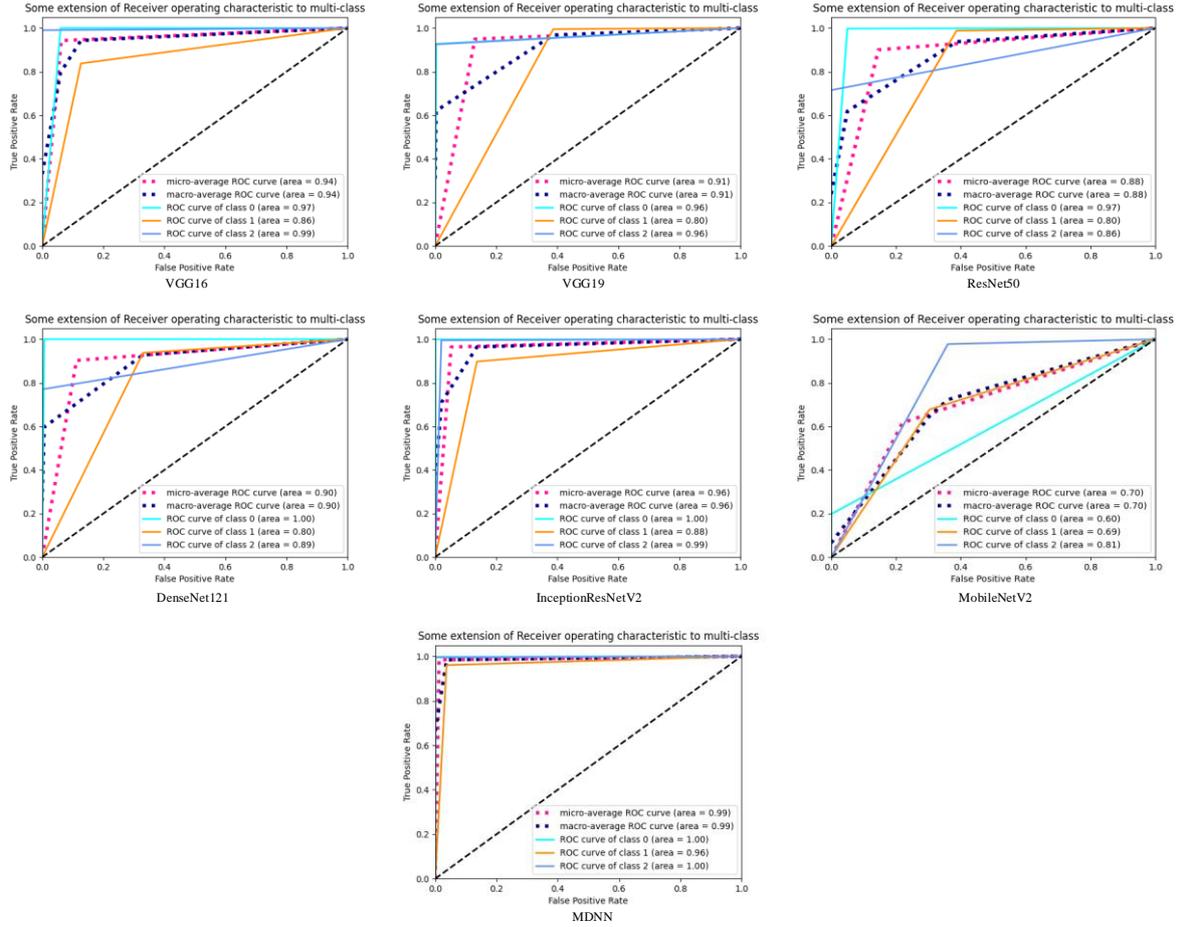

Fig. 5. ROC curves of VGG16, VGG19, ResNet50, DenseNet121, InceptionResNetV2, MobileNet50 and MDNN model.

## B. Result analysis

In this experiment, we use the DNN models which have won in the image recognition competition in recent years, mainly VGG16, VGG19, ResNet50, MobileNetV2, DenseNet121, InceptionResNetV2 were investigated and compared with our model. Experiments have proved that these models perform well in the classification task of the transfer learning. However, due to the different depth and structure of the network algorithm, the classification effect of the trained models on Raman spectrum scale map of mixture is also different. So it is necessary to compare and discuss these models. After each epoch, the order of data and tags is disordered again. Secondly, we use the Early-stopping function to terminate the training and save the trained algorithm model when the loss value does not change on two epochs. In this paper, all models were trained and tested on Tensorflow.2.3-GPU.

Seven trained DNN models are investigated in the experiment. We put the test data set (a total of 700 Raman spectrum scale maps) into the models to test. The evaluation parameters of each algorithm are plotted in Fig. 4. The Hamming loss of DMNN is 0.0140 which is smaller than those of VGG16, VGG19, ResNet50, DenseNet121, InceptionResNetV2 and MobileNetV2 which are 0.0455, 0.0707, 0.1055, 0.0898, 0.0288, and 0.2993 respectively. The ranking loss, coverage, one error and average precision of the MDNN model are 0.04, 0.062, 0.0257 and 0.0367, respectively. These parameters are smaller than those of the other models. F1 macro averaging and F1 micro averaging of the DMNN model are higher than those of others. Base on these indicators, our MDNN model performs best among other six DNN models.

## IV. DISCUSSION

### A. ROC analysis

ROC curve is a comprehensive indicator reflecting the continuous variables of sensitivity and specificity and reveals the relationship between sensitivity and specificity by composition method [49]. It calculates a series of sensitivities and specificities by setting different thresholds for continuous variables [49]. The greater the area under the curve, the higher the accuracy of diagnosis. On the ROC curve, the point closest to the left above the coordinate map was the critical value of sensitivity and specificity. False-positive rate (FPR) on the horizontal axis indicates that the larger the FPR value is, the more negative classes are predicted in the positive class. The true positive rate (TPR) on the vertical axis indicates that the larger the value of TPR, the more actual positive classes in the predicted positive class.

We discuss the relationship between each model for better sensitivity and specificity. From the result part, we find that MDNN model is much better than other existing models in the actual test. The ROC curves of VGG16, VGG19, ResNet50, DenseNet121, InceptionResNetV2, MobileNetV2 and MDNN model are shown in Fig. 5. Under the F1 macro averaging and F1 micro averaging indicators, it is easy to find that the ROC measurement values of ResNet50 and MobileNetV2 models are all lower than 90%. However, the accuracy of VGG16, VGG19, DenseNet121, InceptionResNetV2 models are all over 90%. However, compared with MDNN model, its ROC measurement value is higher than other models. The ROC of each label was predicted by the model. Vgg16, VGG19, DenseNet121 and InceptionResNetV2 models tend to detect the first or the third category when predicting mixtures, while the ROC values classified on the second label are 0.86, 0.80, 0.80, 0.80, 0.88, 0.69, respectively. The error obtained from the MDNN model is only 4% in the second label while other substances can be identified accurately. Compared with other models, MDNN model provides the best performance.

### B. Detection efficiency analysis

In this subsection, the detection times of the proposed model and other compared models are reported. We prepared 700 different kinds of moisy Raman spectra under 20-30 dB SNR. It is shown in Fig. 6 that the time required for the detection of MDNN is only 5.3132s. This is faster than those of VGG16, VGG19, ResNet50, DenseNet121, InceptionResNetV2 and MobileNetV2 which are 7.1245, 7.5046, 8.6300, 12.3294, 16.1131 and 6.6451, respectively. The memory required by MDNN is only 74.5MB, which is small enough to implement a portable Raman spectroscopy solution.

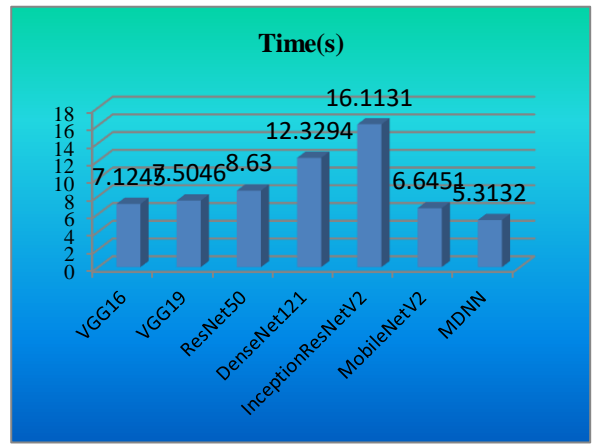

Fig. 6. VGG16, VGG19, ResNet50, DenseNet121, InceptionResNetV2, MobileNet50 and MDNN models were used to test 700 Raman spectra.

## V. CONCLUSION

In this paper, a deep learning algorithm model for multi-label classification and a unified scheme for classifying multi-label mixtures are proposed. In data preprocessing, we avoid the error caused by the baseline correction algorithm and use CWT to extract all the molecular information and noise information of the original Raman spectrum. Secondly, data augmentation is used to improve the imbalance and diversity of training data. In the training process, our MDNN model avoids overfitting and successfully surpasses VGG16, VGG19, ResNet50, DenseNet121, InceptionResNetV2, MobileNet50 in Hamming loss, one error, coverage, ranking loss, average precision, F1 macro averaging and F1 micro averaging. In the ROC index, the measurement value of MDNN model in detecting the first and third kinds of substances is basically similar, but the measurement value of detecting the second kind of substances is significantly higher than other models. Therefore, our model is better than other models. In terms of detection time, our proposed model predicts the Raman spectra of 700 mixtures at 5.3132 seconds, which is much faster than the detection speed of other models. This scheme is of great significance for the detection of mixtures of classified chemicals and paves the way for the combination of Raman spectroscopy and articifical intelligence technology.

**Liangrui Pan** was born in Anhui, China, in 1997. In 2019, he obtained a bachelor's degree from Anhui Polytechnic University. He is pursuing a master's degree in electrical engineering at Prince Songkla University in Thailand in 2019 and is a Member of IEEE and a member of the Chinese Society of Electrical Engineering. His research interests are machine learning, deep learning, and pattern recognition.

**Pronthep Pipitsunthonsan** received a bachelor's degree from Prince of Songkla University in 2010 and a master's degree in 2017. He is currently pursuing a doctorate in computer engineering. Since 2015, he has worked as a programmer at GISTDA. His research interests are deep learning and big data.

**Chalongrat Daengngam** received a B.S. in Physics from Prince of Songkla University, Songkhla, Thailand, in 2005, a M.Sc. in Nanoelectronics & Nanomechanics from University of Leeds, UK in 2006, and a Ph.D. in Physics from Virginia Tech, USA in 2012. Currently, he is working as an assistant professor in the Department of Physics, Faculty of Science, Prince of Songkla University. His research interests involve nonlinear optical properties of nanomaterials, photonics, and standoff Raman spectroscopy.

**Mitchai Chongcheawchamnan** (SM'98) was born in Bangkok, Thailand. He received a B.Eng. degree in telecommunication from the King Mongkut's Institute of Technology Ladkrabang, Bangkok, in 1992, a M.Sc. degree in communication and signal processing from Imperial College, London, U.K., in 1995, and a Ph.D. degree in electrical engineering from the University of Surrey, Guildford, U.K., in 2001. He joined Mahanakorn University of Technology, Bangkok, as a Lecturer, in 1992. In 2008, he joined the Faculty of Engineering, Prince of Songkla University, Songkhla, Thailand, as an Associate Professor. His current research interests include deep learning algorithm and big data applied for agricultural applications and smart cities.